\documentclass[prl,aps,twocolumn,showpacs,floatfix,superscriptaddress]{revtex4}
\usepackage{amsmath,latexsym,amssymb}
\usepackage{graphicx}

\begin{document}

\title{Exciton-polaritons in transition-metal dichalcogenides and their
direct excitation via energy transfer}
\author{Yuri N. Gartstein}
\email{yuri.gartstein@utdallas.edu}
\affiliation{Department of Physics, The University of Texas at Dallas, Richardson, Texas,
75080, USA}
\author{Xiao Li}
\affiliation{Department of Physics, The University of Texas at Austin, Austin, Texas
78712, USA}
\affiliation{Department of Physics, The University of Texas at Dallas, Richardson, Texas,
75080, USA}
\author{Chuanwei Zhang}
\email{chuanwei.zhang@utdallas.edu}
\affiliation{Department of Physics, The University of Texas at Dallas, Richardson, Texas,
75080, USA}

\begin{abstract}
Excitons, composite electron-hole quasiparticles, are known to play an
important role in optoelectronic phenomena in many semiconducting materials.
Recent experiments and theory indicate that the band-gap optics of the newly
discovered monolayer transition-metal dichalcogenides (TMDs) is dominated by
tightly bound valley excitons. The strong interaction of excitons with
long-range electromagnetic fields in these 2D systems can significantly
affect their intrinsic properties. Here, we develop a semi-classical
framework for intrinsic exciton-polaritons in monolayer TMDs that treats
their dispersion and radiative decay on the same footing and can incorporate
effects of the dielectric environment. It is demonstrated how both inter-
and intra-valley long-range interactions influence the dispersion and decay
of the polaritonic eigenstates. We also show that exciton-polaritons can be
efficiently excited via resonance energy transfer from quantum emitters such
as quantum dots, which may be useful for various applications.
\end{abstract}

\maketitle

\textit{Introduction.} Monolayer molybdenum disulfide MoS$_{2}$ and other
group-VI transition-metal dichalcogenides (TMDs) are novel two-dimensional
(2D) semiconductor systems whose electronic and optical properties attract a
great deal of attention \cite%
{Geim2013,Xu2014,Mak2010,Xiao2012,Zeng2012,Mak2012,Cao2012,Jones2013,Zhang2014,Ross2013,Mak2013,Feng2012,Lagarde2014,Jones2014,Mak2014,Baugher2013,Li2013,Chu2014,Aivazian2014,Wu2013,Qiu2013,Yu2014,Liu2013,Berkelbach2013,Berghauser2014,Ramasubramaniam2012,Wu2015}
. One of their prominently discussed features is the opportunity to
manipulate the valley degree of freedom, including by optical means due to
opposite-handed circular polarizations of the interband transitions in the
two valleys \cite{Xiao2012}. A growing experimental and theoretical evidence
\cite%
{Xu2014,Mak2010,Xiao2012,Zeng2012,Mak2012,Cao2012,Jones2013,Ross2013,Jones2014,Aivazian2014,Qiu2013,Wu2013,Yu2014,Liu2013,Berkelbach2013,Berghauser2014,Ramasubramaniam2012,Wu2015}
indicates that the band-gap optical properties of monolayer TMDs are
dominated by relatively tightly bound electron-hole pairs, excitons, with
binding energies substantially larger than in the majority of conventional
inorganic semiconductor quantum wells \cite{Basu1997,Haug2004}. The
corresponding 2D exciton physics in TMDs may therefore reflect generally
stronger interactions of excitons with macroscopic electric fields and light.

An important fundamental issue is the nature of intrinsic excitonic
eigenstates and their energy-momentum (dispersion) relationships. It was
shown recently \cite{Yu2014} that the long-range exchange Coulomb
interaction mixes individual valley excitons to establish excitons with the
longitudinal ($L$) and transverse ($T$) polarizations as normal system
modes, similarly to quantum well excitons \cite{Andreani1990,Girlanda1998}.
The resulting exciton spectrum, as a function of its center-of-mass in-plane
wave-vector $\mathbf{k}=(k_{x},k_{y})=k\,(\cos \theta ,\sin \theta )$, was
found to exhibit a specific Dirac-cone-like behavior at low momenta $\hbar k$%
, in particular within the light cone, $k<q=\omega /c$. As we perform in
this Letter a semi-classical analysis of the interaction of monolayer
excitons with long-range electromagnetic fields, it will be shown however
that (1) the dispersion of $L$- and $T$-excitons is affected via \textit{both%
} inter- and intra-valley processes leading to the overall generic behavior
characteristic of the 2D excitons \cite%
{Andreani1990,Girlanda1998,Agranovich1984,Agranovich2009} without a
Dirac-cone feature; (2) Moreover, also in a general fashion, the intrinsic
behavior within and in the vicinity of the light cone is that of the 2D
exciton-polaritons as determined by the full electromagnetic (rather than
just electrostatic) mixing of valley excitons taking account of the
retardation effects \cite{Haug2004,Agranovich2009}. (In the electrostatic
limit the obtained dispersion reproduces the exciton spectrum that we derive
from inter- and intra-valley exchange Coulomb interactions). General
features of the intrinsic exciton-polaritons are clearly accentuated in the
macroscopic electrodynamics framework, which allows for a straightforward
generalization of the analysis of the polariton dispersion and radiative
decay in free-standing monolayer TMDs to monolayers at the interface between
different media. We also use that framework to illustrate the possibility of
very efficient direct excitation of monolayer polaritons by energy transfer
from proximal electric-dipole emitters such as quantum dots, which may be
useful for various applications.

\textit{Exciton-polaritons from effective dipole-dipole interactions.} In a
common basic description of optically active lowest energy excitons in
monolayer TMDs, they arise as a result of the direct Coulomb attraction
between an electron and a hole in the same valley. Due to the strong
spin-orbit coupling in TMDs, the electron-hole spin composition is
associated with the valley \cite{Xiao2012} and therefore will be omitted. In
the valley-based picture there are then two exciton species (index $\alpha
=1,2$) corresponding to two different valleys, both having the same energy
dispersion
\begin{equation}
E_{0}(\mathbf{k})=\hbar \omega _{0}(k)=\hbar \omega _{0}+\hbar
^{2}k^{2}/2M_{0},  \label{bare}
\end{equation}%
where $\hbar \omega _{0}$ is typically somewhat below 2 eV and the exciton
mass $M_{0}$ is close to the free electron's $m_{e}$ \cite%
{Jones2013,Mak2013,Ross2013,Yu2014}. The parabolic kinetic energy in Eq.~(%
\ref{bare}) with $M_{0}$ being the sum of electron and hole effective masses
signifies the fact that the exciton propagation in space is enabled by the
simultaneous motion of the electron and hole. The inclusion of the standard
exchange Coulomb interaction results in another mechanism of the exciton
propagation and concomitant modifications of its dispersion, which can be
seen as the electric field of the exciton annihilated at one spatial point
creating the exciton at another point. Such a long-range process is
well-known to correspond to the electrostatic dipole-dipole coupling, which
is the major exciton transfer mechanism in molecular systems \cite%
{Agranovich2009}. This physically attractive semiclassical real-space
picture can then be readily extended to the dipole-dipole coupling mediated
by the full electromagnetic interactions with the retardation effects in
place. The resulting intrinsic excitations, exciton-polaritons, would thus
take into account the exciton-light interaction absent in the picture of
Coulomb excitons derived with electrostatic interactions alone \cite%
{Haug2004,Agranovich1984,Agranovich2009}.

The electromagnetic interactions in monolayer TMDs may be classified as
intra-valley ($\alpha =\beta $) and inter-valley ($\alpha \neq \beta $)
couplings, in some analogy with bipartite lattices and molecular systems
with two molecules per unit cell \cite{Agranovich2009}. Correspondingly, the
exciton self-energy correction due to the interactions is a $2\times 2$
matrix $\boldsymbol{\Sigma }$. In the continuum description of 2D
Wannier-Mott excitons, it can be written in the form of
\begin{equation}
\Sigma _{\alpha \beta }(E,\mathbf{k})=\int d\mathbf{r}\,v_{\alpha \beta }(%
\mathbf{r})\,e^{i\mathbf{k}\cdot \mathbf{r}},  \label{Sig}
\end{equation}%
as a function of energy $E=\hbar \omega =\hbar cq$ and wave-vector $\mathbf{k%
}$ variables. Here $v_{\alpha \beta }(\mathbf{r})=|\psi (0)|^{2}\,V_{\alpha
\beta }(\mathbf{r})$ is the 2D energy density determined by the probability $%
|\psi (0)|^{2}=8/(\pi a_{B}^{2})$ of finding the electron and hole of the
exciton at the same spatial point \cite{Haug2004} and the interaction matrix
elements $V_{\alpha \beta }(\mathbf{r})$ dependent on the relative
two-dimensional position $\mathbf{r}=r\hat{\mathbf{r}}$. From \emph{Ab initio%
} calculations \cite{Feng2012,Qiu2013,Ross2013,Berkelbach2013}, the exciton
Bohr radius in monolayer TMDs is estimated as $a_{B}\sim 1$ nm. For the
free-standing monolayers in vacuum the long-range part of the interaction
\begin{eqnarray}
\!\!\!\!\!V_{\alpha \beta }(\mathbf{r}) &=&\frac{e^{iqr}}{r}\left\{ q^{2}%
\left[ (\hat{\mathbf{r}}\cdot \mathbf{d}_{\alpha }^{f0})(\hat{\mathbf{r}}%
\cdot \mathbf{d}_{\beta }^{0\!f})-\mathbf{d}_{\alpha }^{f0}\cdot \mathbf{d}%
_{\beta }^{0\!f}\right] \right.  \notag \\
&\!\!\!\!\!+&\!\!\!\!\!\left. \left( \frac{1}{r^{2}}-\frac{iq}{r}\right) %
\left[ \mathbf{d}_{\alpha }^{f0}\cdot \mathbf{d}_{\beta }^{0\!f}-3(\hat{%
\mathbf{r}}\cdot \mathbf{d}_{\alpha }^{f0})(\hat{\mathbf{r}}\cdot \mathbf{d}%
_{\beta }^{0\!f})\right] \right\}  \label{V12}
\end{eqnarray}%
as arising from the full electromagnetic dipole-dipole coupling (in Gaussian
units) \cite{Jackson1975}. The familiar electrostatic limit corresponds to
the light vacuum wave number $q=\omega /c\rightarrow 0$ (speed of light $%
c\rightarrow \infty $). The interacting $\alpha $ and $\beta $ species in
Eq.~(\ref{V12}) are represented by the corresponding interband dipole
transition matrix elements for creation, $\mathbf{d}_{\alpha }^{f0}$, and
annihilation, $\mathbf{d}_{\beta }^{0\!f}$, of electron-hole pairs. In the
electrostatic limit, the self-energy corrections (\ref{Sig}) would be
real-valued and functions of wave vector $\mathbf{k}$ only. With the
retarded electric fields, however, $\Sigma _{\alpha \beta }$ are functions
of both energy and wave-vector variables and generally complex-valued; they
would therefore determine both the exciton dispersion and the decay width
(decay into photons) in a self-consistent calculation. Calculations with the
retarded fields are equivalent \cite{Gartstein2007} to calculations of
exciton-polaritons as arising from the interaction of Coulomb excitons with
transverse photons \cite{Haug2004,Agranovich2009}.

\begin{figure}[tbp]
\includegraphics[scale=0.52]{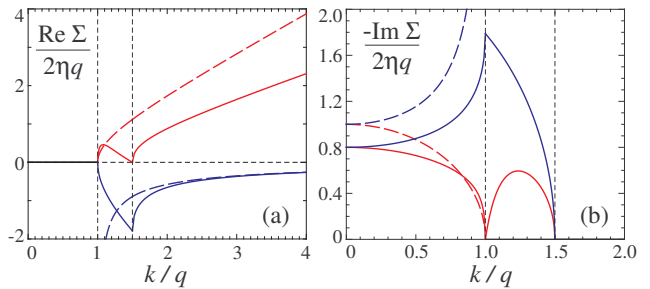}
\caption{Functional behavior of the (a) real and (b) imaginary parts of the
intrinsic exciton-polariton self-energy as per Eqs.~(\protect\ref{LT1}) and (%
\protect\ref{LT2}). Red color lines are used for the longitudinal ($L$) and
blue for the transverse ($T$) excitons. Shown with dashed lines are the
results for a free-standing monolayer in vacuum, and solid lines are for the
monolayer on a glass-like substrate, see text.}
\label{Fig:Exciton}
\end{figure}

The in-plane interband dipole transition matrix elements $\mathbf{d}%
_{\alpha}^{f0}=\left(d_x^{\,(\alpha)},id_y^{\,(\alpha)}\right) $ have a
common form for both valleys but with the opposite handedness of their
circular polarization:
\begin{equation}  \label{dip}
d_{x}^{(1)}=d_{y}^{(1)}=d_{x}^{(2)}=-d_{y}^{(2)}=\mathcal{D},
\end{equation}
where we chose $\mathcal{D}$ as a real positive quantity. From a two-band
monolayer TMD model \cite{Xiao2012}, for instance, the dipole transition
moment would be $\mathcal{D}=eat/E_{g}$, where $e$ is the fundamental
charge, $a$ the lattice structure constant ($\simeq 3.19$ \AA ), $t$ the
nearest neighbor hopping integral ($\simeq 1.10$ eV) and $E_{g}$ the energy
gap between the conduction and valence bands that the excitons come from. A
calculation based on the three-band model \cite{Liu2013} gives the same
form, but with a slightly different value of $t$.

The difference between the transition dipole moments (\ref{dip}) in the two
valleys translates into the difference of inter- and intra-valley couplings (%
\ref{V12}). A direct calculation of integrals in Eq.~(\ref{Sig}) then leads
to the following $\mathbf{k}$-dependence of the self-energy matrix \cite%
{note1}:
\begin{equation}
\boldsymbol{\Sigma }(E,\mathbf{k})=\left(
\begin{array}{cc}
J_{0} & \exp (-2i\theta )\,J_{1} \\
\exp (2i\theta )\,J_{1} & J_{0}%
\end{array}%
\right) ,  \label{Sig1}
\end{equation}%
where the intra-valley component $J_{0}=-i\eta (\sqrt{q^{2}-k^{2}}+q^{2}/%
\sqrt{q^{2}-k^{2}})$ and the inter-valley component $J_{1}=i\eta k^{2}/\sqrt{%
q^{2}-k^{2}}$ feature the same magnitude scale $\eta =2\pi |\psi (0)|^{2}%
\mathcal{D}^{2}$. The functional form of these components illustrates the
great qualitative distinction resulting from the retarded interactions: the
obtained corrections $J_{0}$ and $J_{1}$ are purely imaginary above the
light line, $k<q$, but become purely real ($\sqrt{q^{2}-k^{2}}\rightarrow i%
\sqrt{k^{2}-q^{2}}$) below the light line, $k>q$. As is known \cite%
{Agranovich1984,Agranovich2009}, this signifies the impossibility for the
intrinsic 2D exciton-polariton with $k>q$ to decay into a photon due to the
momentum conservation. Exciton-polaritons with $k<q$, on the other hand,
exhibit the radiative width due to such a decay. Moreover this width is
greatly enhanced in comparison with localized emitters \cite{Agranovich2009}%
. It is also instructive to look at the electrostatic limit ($q\rightarrow 0$%
) of these corrections: the inter-valley component then becomes $J_{1}=\eta
k $, precisely the result derived in Ref.~\cite{Yu2014}. Importantly, the
intra-valley component $J_{0}$ becomes equal to $J_{1}$ in this limit, which
is also confirmed in our independent calculations with the exchange Coulomb
interaction.

Because of the inter-valley coupling $J_{1}$, the valley excitations are
clearly not the eigenstates of the system. Instead, the eigenstates are
their linear combinations $\Psi _{\pm }=2^{-1/2}\left(
\begin{array}{c}
1 \\
\pm \exp (2i\theta )%
\end{array}%
\right) $ that diagonalize matrix $\boldsymbol{\Sigma }$ and yield the
corresponding self-energy corrections as
\begin{equation}
\Sigma _{\pm }=J_{0}\pm J_{1}=-2i\eta \times \left\{
\begin{array}{ll}
\sqrt{q^{2}-k^{2}},\ \ \ \ \ \ (L), &  \\
q^{2}/\sqrt{q^{2}-k^{2}},\ \ (T). &
\end{array}%
\right.  \label{LT1}
\end{equation}%
These eigenstates have, respectively, longitudinal, for $\Psi _{+}$, and
transverse, for $\Psi _{-}$, polarizations with respect to polariton
wave-vector $\mathbf{k}$ \cite{Yu2014}: for the propagation along the $x$%
-axis, e.g., Eq.~(\ref{dip}) shows that $\Psi _{+}$ combination corresponds
to the transition moment $\propto (\mathcal{D},+i\mathcal{D})+(\mathcal{D},-i%
\mathcal{D})$ along $x$ while $\Psi _{-}$ to the moment $\propto (\mathcal{D}%
,+i\mathcal{D})-(\mathcal{D},-i\mathcal{D})$ along $y$. The functional forms
of self-energy corrections (\ref{LT1}) for the free-standing monolayer are
shown in Fig.~\ref{Fig:Exciton} by dashed lines. It is clear that, in the
electrostatic limit, only the $L$-excitons would acquire the additional term
$\Sigma _{+}=2\eta k$ in their dispersion due to long-range exchange
interactions, while the dispersion of $T$-excitons would remain unchanged ($%
\Sigma _{-}=0$) as Eq.~(\ref{bare}). There is no Dirac-cone-like behavior
even in the electrostatic limit. In the valley-centric basis these
conclusions thus follow from the simultaneous account of both inter- and
intra-valley processes as opposed to the inter-valley coupling alone \cite%
{Yu2014}.

\textit{Exciton-polaritons from macroscopic Maxwell equations.} The obtained
results for polaritonic eigenstates are in agreement with the picture known
for quantum wells \cite{Andreani1990} and reflect the fact that the
opposite-handedness in-plane susceptibilities $\boldsymbol{\chi}_1$ and $%
\boldsymbol{\chi}_2$ of the individual valleys just add up in the overall
\textit{isotropic} electrodynamic response of the monolayer. The latter is
then characterized by the scalar susceptibility $\chi$ defining the
monolayer 2D current density $\mathbf{j}=-4\pi i\omega\chi \mathbf{E}$
induced by the in-plane electric field $\mathbf{E}$. It is this current
density that enters the boundary conditions for the macroscopic Maxwell
equations determining the effects of long-range fields on the system
excitations \cite{Haug2004,Agranovich2009}. For well-separated excitonic
states (\ref{bare}), e.g., the 2D scalar susceptibility acquires a familiar
single-oscillator form
\begin{equation}  \label{chi}
\chi(\omega,k)=\chi_0 + A/\left(\omega_0^2(k)-\omega^2-2i\gamma\omega\right),
\end{equation}
where $\chi_0$ is the background term due to higher-frequency transitions
and $\gamma$ the phenomenological dissipation parameter. A many-oscillator
form could be used to include even more specifics for different TMD
monolayers.

\begin{figure}[tbp]
\includegraphics[scale=0.45]{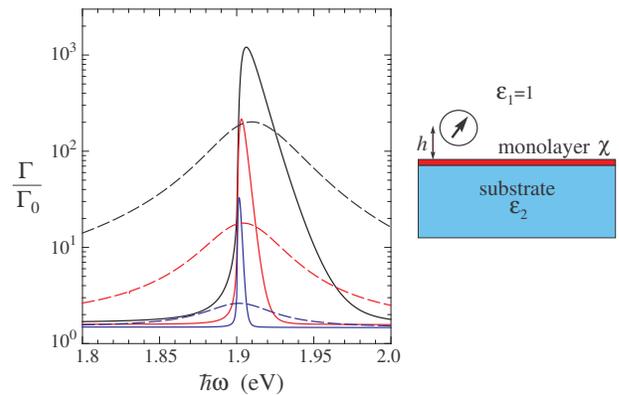}
\caption{Acceleration of the decay rate $\Gamma $ of a randomly oriented
electric dipole emitter such as quantum dots in the vicinity of the
monolayer on a glass-like substrate as compared to the spontaneous decay
rate $\Gamma _{0}$ in vacuum, Eq.~(\protect\ref{Ga}). The data are shown as
a function of the emitter frequency $\protect\omega $ and for different
distances $h$ to the interface: 5 nm (black lines), 10 nm (red) and 20 nm
(blue). Solid lines display results for the nearly dissipationless excitons (%
$\protect\gamma =0.1$ meV), dashed lines are for the dissipation parameter $%
\protect\gamma =25$ meV. These illustrative calculations were done with the
model parameters $4\protect\pi \protect\chi _{0}=2$ nm, $4\protect\pi \hbar
^{2}A=0.25$ eV$^{2}\cdot $nm in Eq.~(\protect\ref{chi}) and $\hbar \protect%
\omega _{0}=1.9$ eV, $M_{0}=m_{e}$ in Eq.~(\protect\ref{bare}).}
\label{Fig:ET}
\end{figure}

The versatile macroscopic framework can be easily applied to various
environments. Here we exemplify this by considering the model of an
infinitesimally thin planar monolayer between two non-magnetic media with
dielectric constants $\varepsilon _{1}$ and $\varepsilon _{2}$. One can
solve for the eigenfrequencies of the system directly. Alternatively, as
also useful for other problems, one looks at the poles of the reflection
coefficients for electromagnetic waves in our sandwich configuration. The 3D
setup involves not only the in-plane components of the fields and wave
vectors but also their $z$-components perpendicular to the planar interface.
With the boundary conditions of the polarizable interface monolayer \cite%
{note2}, one easily derives the reflection coefficient amplitudes for $p$-
and $s$-polarized waves as $(\omega ,k)$-dependent
\begin{equation}
r^{(p)}=\left( \frac{\varepsilon _{2}}{k_{z2}}-\frac{\varepsilon _{1}}{k_{z1}%
}-4\pi i\chi \right) \left( \frac{\varepsilon _{2}}{k_{z2}}+\frac{%
\varepsilon _{1}}{k_{z1}}-4\pi i\chi \right) ^{-1}  \label{rp}
\end{equation}%
and
\begin{equation}
r^{(s)}=\left( k_{z1}-k_{z2}+4\pi iq^{2}\chi \right) \left(
k_{z1}+k_{z2}-4\pi iq^{2}\chi \right) ^{-1}.  \label{rs}
\end{equation}%
The normal components $k_{zi}=(\varepsilon _{i}q^{2}-k^{2})^{1/2}$ of the
waves in the respective media appear, as usual \cite{Novotny2006}, related
to the in-plane wave number $k$ for given frequency $\omega =cq$. The
in-plane component of the electric field in a $p$-polarized wave is along
the in-plane wave vector $\mathbf{k}$ whereas in an $s$-polarized wave they
are perpendicular to each other. Hence, the poles of Eq.~(\ref{rp})
determine the dispersion and decay of the $L$-polaritons while poles of Eq.~(%
\ref{rs}) those of the $T$-polaritons. Writing the pole equations in the
form of
\begin{equation}
-\frac{1}{\chi }=-4\pi i\times \left\{
\begin{array}{ll}
\left( \varepsilon _{1}/k_{z1}+\varepsilon _{2}/k_{z2}\right) ^{-1},\ \ (L),
&  \\
q^{2}/(k_{z1}+k_{z2}),\ \ \ \ \ \ \ \ \ \ (T), &
\end{array}%
\right.  \label{LT2}
\end{equation}%
one recognizes that the functional dependence in the r.~h.~s. of Eq.~(\ref%
{LT2}) reduces to that in Eq.~(\ref{LT1}) for the free-standing monolayer.
It is also clear that with the negligible dissipation and screening due to
higher-frequency transitions, $-1/\chi \simeq 2\omega _{0}(\omega -\omega
_{0}(k))/A$ becomes just proportional to the self-energy corrections in the
vicinity of the resonance, $\omega \simeq \omega _{0}$. The derivation with
the effective dipole-dipole interactions thus fully agrees with the
macroscopic electrodynamics result, and we obtain $A=2\omega _{0}\eta /\pi
\hbar $ by comparing Eqs.~(\ref{LT1}) and (\ref{LT2}). Note that the
screening by the surrounding can substantially affect the exciton radius and
binding energy reducing thus \textquotedblleft the
strength\textquotedblright\ $A$ of the resonance (\ref{chi}); the
calculations of exciton binding are outside of our scope.

Figure \ref{Fig:Exciton} illustrates the differences in the behavior of
self-energy corrections for two systems: the symmetric sandwich with $%
\varepsilon_1=\varepsilon_2=1$ (a free standing monolayer) and and the
asymmetric sandwich with $\varepsilon_1=1$ and $\varepsilon_2=2.25$
(glass-like substrate), as follows from Eq.~(\ref{LT2}). It is transparent
that in the symmetric case, the $T$-polariton branch exhibits splitting at
the light cone, $k\rightarrow q$, as consistent with the divergence of the
radiative decay rate (\ref{LT1}) at $k\rightarrow q-0$. Qualitatively
differently for the asymmetric sandwich, there is no divergence in the decay
rate and the dispersion of the $T$-polariton branch is continuous. Figure %
\ref{Fig:Exciton}(b) also shows the extension of the region of the radiative
decay: polaritons with $q < k < \sqrt{\varepsilon_2}q$ can now decay into
photons that propagate only in the substrate. Beyond that region, however,
the intrinsic (without scattering effects) polaritons become non-radiative
and, conversely, cannot be directly excited by plane-wave photons incident
on the monolayer from the infinity, the situation similar to the excitation
of surface waves like surface plasmons \cite%
{Agranovich1984,Agranovich2009,Novotny2006}.

\textit{Application to energy transfer.} The direct excitation of
non-radiative modes is however known to be possible by the near
electromagnetic fields. This may be accomplished, e.g., with special optical
geometry setups \cite{Agranovich1984,Agranovich2009,Novotny2006} or via
resonance energy transfer (ET) from proximal quantum emitters such as
molecules or quantum dots \cite{Novotny2006,Philpott1975,Chance1978}. We
note that ET from 0D emitters to (quasi) 2D absorbers has been of increasing
theoretical and experimental interest, including organic and inorganic
semiconductors \cite{Gordon2013}, graphene \cite{Gaudreau2013} and MoS$_2$
\cite{Prins2014} (and references therein). ET provides a new decay channel
and is thus manifested in the acceleration of the observed emitter's
photoluminescence decay.

Using the macroscopic electrodynamics formalism developed for such
applications \cite{Novotny2006,Chance1978}, the decay rate $\Gamma $ of the
randomly-oriented electric-dipole emitter in the medium with dielectric
constant $\varepsilon _{1}$ at distance $h$ from the planar interface is
derived \cite{Novotny2006} as
\begin{eqnarray}
\Gamma /\Gamma _{1}=1 &+&(1/2k_{1}^{3})\,\mathrm{Re}\int_{0}^{\infty
}(k\,dk/k_{z1})\,e^{2ik_{z1}h}  \notag  \label{Ga} \\
&\times &\left( (2k^{2}-k_{1}^{2})r^{(p)}+k_{1}^{2}r^{(s)}\right) ,
\end{eqnarray}%
where $\Gamma _{1}$ is the spontaneous decay rate in the uniform medium and $%
k_{1}^{2}=\varepsilon _{1}q^{2}$. The integration in Eq.~(\ref{Ga}) over all
values of the in-plane wave-vectors $k$ signifies that the near-fields of
the emitter are included. Figure \ref{Fig:ET} illustrates the results
following from Eq.~(\ref{Ga}) with the reflection coefficients (\ref{rp})
and (\ref{rs}) for the considered geometry. The analysis shows that a very
large effect observed in Figure \ref{Fig:ET} is predominantly due to the
excitation of $L$-polaritons (poles of $r^{(p)}$), similar to a very
efficient excitation of surface plasmons by ET \cite{Philpott1975,Chance1978}%
; the dispersion of the peak with distance $h$ is also clearly seen. For the
computational illustration here we used a more moderate value of the
resonance strength $A$ in Eq. (\ref{chi}), and its larger values would lead
to even faster ET into the monolayer. Figure \ref{Fig:ET} also shows how the
narrower polariton peaks are spread by exciton damping ($\gamma $)
processes. Excitation of TMD monolayer polaritons via high efficiency ET
from neighboring emitters might be attractive for various optoelectronic
applications \cite{Prins2014,Agranovich2011}. Its experimental studies for
different emitter frequencies and distances could be used for quantification
of polariton properties.

\textit{Discussion.} While our discussion in this paper has been focussed on
the intrinsic polaritons, it is recognized that various scattering processes
due to polariton interactions with phonons and defects can significantly
affect the observable optical properties of realistic TMDs, and this appears
as an important topic for future studies. Just as with the quantum well
polaritons, e.g., their thermalization may lead to a specific temperature
dependence of the radiative lifetime. With the substantial contribution to
polariton dispersion in TMDs from the long-range interactions we discussed,
that temperature dependence may deviate from the dependence resulting from
the purely parabolic exciton dispersion \cite{Feldman1987}.

Finally, we estimate the experimentally relevant parameters for MoS$_2$.
Following Ref.~\cite{Yu2014}, $\eta $ would be estimated as $\sim 0.75$ eV$%
\cdot $\AA , resulting in substantial contributions to the exciton
dispersion. With this estimate, the energy unit in Fig.~\ref{Fig:Exciton} $%
2\eta q\sim 1.5$ meV for $q= \omega/c \sim 0.01$ nm$^{-1}$. Likewise, one
would obtain $4\pi \hbar^{2}A\sim 1.1$ eV$^{2}\cdot$nm for the unscreened
value of the parameter $A$ in Eq.~(\ref{chi}).

\begin{acknowledgments}
\textbf{Acknowledgements:} YNG is grateful to NSF for the support provided
through the DMR-1207123 grant. XL and CZ are supported by ARO
(W911NF-12-1-0334) and AFOSR (FA9550-13-1-0045).
\end{acknowledgments}


\begin{thebibliography}{99}
\bibitem{note1} The integral (2) with interaction (3) would formally diverge
due to the short-distance $\propto 1/r^{3}$ behavior of the electrostatic
dipole-dipole coupling. Of course, at short distances the point-dipole
description does not apply, whereas the actual short-range Coulomb
interactions would be determined by the detailed behavior of electron wave
functions. The divergent electrostatic term needs therefore to be separated
in the calculation. The separated term represents a constant overall
renormalization shift $\Sigma _{0}$ of the exciton energies due to
short-range interactions that we are not interested in.
\bibitem{note2} We restrict our attention here to the layer with the inplane
polarizability only. With account of the polarizability perpendicular to the
layer, the expression for the reflection coefficient $r^{(p)}$ is more
involved than Eq. (8).
\end{thebibliography}

\end{document}